\documentclass[useAMS,usenatbib,referee,usegraphicx]{mn2e}
\usepackage{enumitem}
\usepackage{graphicx}
\usepackage{caption}
\usepackage{amssymb}
\usepackage{amsmath}
\usepackage{multirow}
\usepackage{mathrsfs}
\usepackage{bbding}
\usepackage{float}
\usepackage{color}
\usepackage{hyperref}
\usepackage[latin9]{inputenc}
\newcommand{\beq}{\begin{equation}}
\newcommand{\eeq}{\end{equation}}
\newcommand{\be}{\begin{equation}} \newcommand{\ee}{\end{equation}}
\newcommand{\bea}{\begin{eqnarray}} \newcommand{\eea}{\end{eqnarray}}

\title[Extracting Spectral Index of Intergalactic Magnetic Field from Radio Polarizations]
      {Extracting Spectral Index of Intergalactic Magnetic Field from Radio Polarizations}
\author[Prabhakar Tiwari and Pankaj Jain]
 {Prabhakar Tiwari \thanks{Technion- Israel Institute of Technology,
32000 Haifa, Israel} and Pankaj Jain \thanks{Department of Physics, Indian
        Institute of Technology, Kanpur - 208016, India} }

\begin{document}
\maketitle
\begin{abstract}
We explain the large scale correlations in  radio polarization in terms of the correlations of
galaxy cluster/supercluster magnetic field. Assuming that the polarization correlations 
closely follow the spatial 
correlations of the background magnetic field we recover the magnetic field spectral 
index as $-2.74\pm 0.04$. 
This remarkably agrees with cluster magnetic
field spectral index obtained 
in cosmological magneto-hydrodynamic simulations. We discuss possible
physical scenarios in which the observed polarization alignment is plausible.

\end{abstract}
\begin{keywords}
polarization, galaxies: high-redshift, galaxies: active
\end{keywords}
\section{Introduction}
The polarization directions from distant radio sources have been observed to 
show an alignment over a distance scale of order 100 Mpc\citep{PTiwari:2013pol}. 
This alignment is seen in the significantly polarized sources in  JVAS/CLASS 
data \citep{Jackson:2007} with polarized flux greater than 1 mJy. 
Such a global alignment of polarization angles is unexpected but not
in conflict with any fundamental principle. The distance scale 100 Mpc 
corresponds to the scale of galaxy superclusters and at such distances it
is not unreasonable that the galaxies may show some correlation with
one another. A similar distance scale also emerges in the study of galaxy 
correlations using Sloan Digital Sky Survey \citep{Eisenstein:2005,Anderson:2014} and
agrees with predictions of the Big Bang cosmological model. 
However a precise physical mechanism which might lead to
an alignment of polarizations is so far not available in the literature. 
Polarization measurements are affected by several instrumental and
observational biases which may also lead to the observed signal.
We discuss some of these in section \ref{sc:data}. 
An important observation is that 
  the alignment is absent  for low polarizations \citep{PTiwari:2013pol}.
 This provides some restriction on any explanation in terms of
instrumental bias.  
 Nevertheless, the issue of contribution due to bias requires a 
more detailed study. Furthermore the significance of the alignment effect is 
found to be approximately three sigmas \citep{PTiwari:2013pol}. As we shall
show in the present paper, the significance is further reduced if we
take the jackknife errors into account. Hence the effect may also be 
simply due to a statistical fluctuation. On the other hand, 
assuming that the 
observations represent a real effect, it is interesting to think of 
possible physical processes 
producing such large scale global radio polarization alignments. 

In this paper we propose a physical process which may 
potentially explain the observed 
alignment. The model is based on two major assumptions: 
\begin{itemize}
\item[1.]
The galaxy jet axis is correlated with the cluster magnetic field. 

\item[2.] The integrated radio polarization is correlated with the jet axis.
\end{itemize}
The galaxy jet axis is assumed to mark the axis of the galactic disk. 
The second assumption is justified by 
several observations \citep{Gabuzda:1994,Lister:2000,Pollack:2002,Helmboldt:2007,Joshi:2007} which indicate that the  
 integrated polarization from such sources 
is predominantly perpendicular or, less frequently, aligned with jet axis. 
Furthermore, the galaxies are known to be statistically aligned over 
large distance
scales, although a proper physical 
understanding of this phenomenon is so far lacking (see e.g. \cite{Kirk:2015,Kiessling:2015}, 
and references therein). 
The cluster magnetic field strength, as observed in cosmological magneto-hydrodynamic simulations,
closely follows the cluster matter density profile outside the core region 
of the cluster 
\citep{Dolag:2002}. 
The power spectrum of the 
magnetic field can be approximated by a power law with an 
exponent $\sim -2.7$\citep{Dolag:2002}. 
Given that the galaxies show some alignment over large distances and the
fact that cluster magnetic field is correlated with the matter
density profile may provide 
some motivation for our first assumption.  
The presumed magnetic field is expected to show some large
scale correlations in real space. This is discussed in more detail
in Section \ref{sc:PMF}. Since, by our assumptions, the integrated radio polarizations
are correlated with the background magnetic field, we expect the
polarizations of different galaxies to be aligned with one another over the
cluster or supercluster distance scale.

The JVAS/CLASS data sources are core-dominated flat spectrum radio source and 
are predominantly quasars and BL Lacs. The 
integrated polarization from these sources is a few percent. 
Here we consider only the significantly polarized sources for which the polarization is greater than
1 mJy. 
Based on the assumptions stated above, our physical mechanism implies that
the integrated radio polarizations are correlated with the
cluster magnetic field. Hence the observed alignment of the radio polarizations contain information about 
correlations of the magnetic field. We use this relationship in 
order to extract the spectral index of the cluster magnetic field, whose
power spectrum is assumed to follow a power law
\citep{Dolag:2002}. We point out that 
the primordial magnetic field 
\citep{Subramanian:2003sh, Seshadri:2005aa,Seshadri:2009sy,Jedamzik:1996wp,Subramanian:1997gi} is also expected to show a power law behaviour, however, in
this case the spectral index is expected to be very different in comparison
to the expectation for the cluster magnetic field \citep{Dolag:2002}. 
We simulate the cluster magnetic field for some assumed value of the 
spectral index. 
The correlations of the magnetic field directions 
at different spatial positions
are assumed to be directly related to the corresponding correlations
of the polarization angles, i.e. the latter provides an unbiased estimate
of the magnetic field correlations. 
We study these correlations by defining a statistic $S_D$
or $S'_D$, as discussed in Section \ref{sc:SD}. By making a fit to the data statistic
we extract the spectral index of the cluster magnetic field.  
The polarization data is likely to have large scatter and we use the 
 jackknife estimate of errors. 
The jackknife errors are found to be large in comparison to the squared
variance of  alignment statistics
of shuffled PAs \citep{PTiwari:2013pol}. In our earlier determination
of the significance of alignment we had used the latter procedure
\citep{PTiwari:2013pol}. If we instead use the jackknife errors we find
that the significance of alignment is reduced.
However, even with jackknife errors, the alignment signal is found
to be good enough to 
sharply constrain the spectral index of magnetic field assuming the model 
presented above. We hope that the situation 
will improve with future Square Kilometre Array (SKA) observations.

The paper is organized as follows. In Section \ref{sc:PMF} we describe the 
magnetic field model and explain its correlations in real space.
We also explain our numerical procedure to generate a full $3D$ realization of 
magnetic field for a particular set of parameters.
In Section \ref{sc:data} we give details of the JVAS/CLASS data and discuss the 
observed alignments. In Section \ref{sc:SD} we review different
statistical measures of alignment used in this paper. 
We describe our procedure in Section \ref{sc:method}. 
We present our results in Section \ref{sc:results} and conclude in Section \ref{sc:conclusion}.

\section{Correlations in Inter-Galactic Magnetic Field }
\label{sc:PMF}
Magnetic field has been observed at all scales in inter-galactic medium. 
However the origin of observed magnetic field is unknown and most likely to be primordial 
\citep{Subramanian:2003sh,Seshadri:2005aa,Seshadri:2009sy,Jedamzik:1996wp,Subramanian:1997gi}.
In our analysis we need to simulate the intergalactic magnetic field. 
On the cluster scale,  
cosmological magneto-hydrodynamic 
simulations suggest that the power spectrum of the magnetic field
 can be modelled as a power law   
with an exponent $\sim-2.7$ \citep{Dolag:2002}.
However on larger distance scales we expect that this exponent may be different. 
Hence we expect that a simple power law may not be valid at all distance scales. 
Here we assume a simple power law form of the power spectrum at all scales with  
an exponent corresponding to the cluster magnetic field. This will correctly
reproduce the magnetic field correlations on the cluster scale, which is the only
scale of interest in our analysis. It will fail at larger distances which are not of
interest in the present work. Hence this failure cannot
affect our results. 

Let $b_i(\vec k)$ represent the magnetic field in Fourier space.
We can express its two point correlations as,
\bea
\label{eq:k_corr}
\left<b^*_i(\vec k\,) b_j(\vec q\,)\right> = \delta_{\vec k,\vec q}\,
P_{ij}(\vec k\,) M(k)
\eea
where  $k=|\vec k|$ and 
$P_{ij}$ is the projection operator given as,
\bea
P_{ij} = \left(\delta_{ij} - {k_i k_j\over k^2}\right) \ ,
\eea

The real space magnetic field can be written as,
\bea
B_i(\vec r) = {1\over V} \sum_k b_i(\vec k) e^{i\vec k\cdot r}
\eea
where $V$ is the volume. We assume a power law dependence of the 
spectral function $M(k)$
\bea
M(k) = Ak^{n_B}, 
\label{eq:mk}
\eea
with the spectral index $n_B>-3$. 
The magnetic field is assumed to be statistically uncorrelated 
in $k$-space. However the field 
is correlated in real space and the nature of correlation is controlled by the index $n_B$.
We can write the real space correlation of the field as the Fourier 
transform of equation (\ref{eq:k_corr}). We obtain,
\beq
\label{eq:r_corr}
\left<B_i(\vec r + \vec r^{\,\prime}) B_j(\vec r)\right> = \frac{1}{V} \int d^3 k e^{i {\boldsymbol k}.{\boldsymbol r^{\prime}}}
P_{ij}(\vec k) M(k) W^2 (k r_G),
\eeq
where $r_G$ is the "galactic" scale taken as 1 Mpc and 
$W$ is a window function of the form, 
\beq
W(x) = \left\{
    \begin{array}{l l}
      1  & x<1 \\
      0  & x>1
    \end{array}\right.
\eeq
This window function fixes the value of the 
 scale $r_G$. 
The magnetic field 
$B(\vec r)$ is assumed to be uniform  
 over the scale r$_G=k_G^{-1}= 1$ Mpc. In equation (\ref{eq:r_corr}) we have also 
taken the continuum limit and replaced
 $\sum_k$  by $\frac{1}{V} \int d^3 k$.
The constant $A$ in equation (\ref{eq:mk}) is equal to 
$V \pi^2 B^{2}_0 \frac{3+n_B}{k^{3+n_B}_G}$. It is 
fixed by demanding $\sum_i \left<B_i(\vec r) B_i(\vec r)\right> = B^{2}_{0}$,
where $B_0$ is the intergalactic magnetic field averaged
 over the distance scale of $r_G=1$ Mpc. 
We expect that for the case of the primordial field, 
$B_0\sim$ nG \citep{Seshadri:2009sy,Yamazaki:2010nf,DeAngelis:2008}, 
although this value plays no role in our analysis. Furthermore
 we add that the radio sources considered in this work are separated from one another by a mean 
distance of tens of Mpc (Fig. \ref{fig:nv_dis}) and, hence, averaging 
magnetic field over 1 Mpc is reasonable.

It is also appropriate to demand a large scale cut-off for the correlations in 
equation (\ref{eq:r_corr}). Here we assume that this cutoff is sufficiently large ($r_{max}>$3Gpc) so that we
can simply set $r_{max}$ to be $\infty$. 
Hence, we set the lower limit of integration in  equation (\ref{eq:r_corr}) 
as $k_{min}=r_{max}^{-1}=0$. 
We emphasize that such large scale correlations are expected within the Big-Bang cosmology. 
The perturbations at the time of inflation have wavelengths larger than the 
comoving 
size of the current observable universe and hence the primordial magnetic field correlations may 
also exist over the horizon scale. 

We numerically generate intergalactic magnetic field in 
$3D$ space using the procedure described in \cite{Agarwal:2012} 
for different values of the spectral index
$n_B$. We consider discretized space consisting of a large number of 
cells or domains of equal size. The magnetic field is assumed to be uniform in each domain. 
We first generate the magnetic field in $k$-space, which is straightforward 
since the corresponding field is uncorrelated. 
It is convenient to use polar coordinates ($k$, $\theta$, $\phi$)
in $k$-space. The projection operator $P_{ij}(\vec k)$ ensures that 
the component of magnetic field along $\vec k$, $b_k$ is zero. The remaining two orthogonal component 
$b_{\theta}$ and $b_{\phi}$ are uncorrelated and therefore we generate these by assuming the Gaussian 
distribution, 
\beq
\label{eq:k_dist}
        f(b_{\theta}({\boldsymbol k}),b_{\phi}({\boldsymbol k})) = N \ {\rm exp} 
\left[-\left(\frac{b_{\theta}^{2}({\boldsymbol k}) + b_{\phi}^{2}({\boldsymbol k})}{2M({\boldsymbol k})}\right) \right],
\eeq
where $N$ is the normalization. The distribution in 
equation (\ref{eq:k_dist}) represents an uncorrelated magnetic field in 
$k$-space. Next, we do a Fourier transform to obtain magnetic field in real 
space.   
 
\section {Data}
\label{sc:data}

We use the catalogue produced by \cite{Jackson:2007}. The catalogue contains 12743 core dominated 
flat spectrum radio sources and provides their angular coordinates and the Stokes $I$, 
$Q$ and $U$ parameters. The observable required for the alignment study is the polarization angle (PA).
A detailed description of the catalogue including the calibration methods is given in \cite{Jackson:2007}. 
We find that there are 4400 sources in catalogue with polarized flux greater than 1 mJy. 
JVAS/CLASS sources are presumably located at very large distances ($z\sim1$), although the 
exact redshift of each source is unknown. A signal of alignment of radio 
polarization angles (PAs) has been found in significantly polarized 
sources, with polarized flux greater than 1 mJy, contained in this catalogue \citep{PTiwari:2013pol}. 
The alignment was seen over the distance scales of order 100 Mpc. At larger distances the authors
\citep{PTiwari:2013pol} do not find a significant  signal of alignment and confirm the earlier null result in
\cite{Joshi:2007}. A recent study of this data with an alternate  statistical measure also 
indicates alignment in data \citep{Shurtleff:2014}. Furthermore alignments have been found even 
at larger distances for QSOs in this data sample \citep{Pelgrims:2015}.

The data may be affected by some instrumental and observational
biases \citep{Joshi:2007,Jackson:2007}. One possibility is the error in the 
removal of residual instrumental 
polarization. This may artificially generate large scale alignments. 
However \cite{PTiwari:2013pol} argue that this must dominate for sources
with low polarizations which do not show any signal of alignment. Hence
it is not possible to attribute the observed alignment to this bias. 
Another possibility is that sources in a small neighbourhood are observed
together within a particular observational run. It is possible that this
could generate alignment in sources within small angular separations, 
as observed in \cite{PTiwari:2013pol}. We cannot rule out such a possibility. 
This issue is best addressed by future more refined observations.

\begin{figure}
\includegraphics[width=3.5in,angle=0]{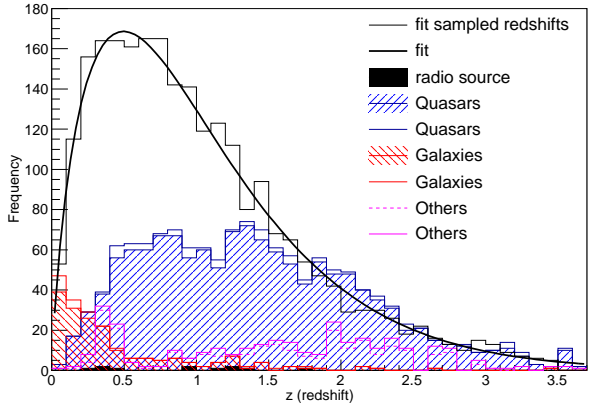}
  \caption{ The redshift distributions of different classes of JVAS/CLASS objects. A total 
of 1921 sources out 
of 4400 are retrieved from the NED database.  
The filled (or dashed line) histogram corresponds to NED object match within $0.1$ arcsec, 
whereas the empty 
(or solid line) corresponds to NED object match within $0.5$ arcsec.  Assuming that the 
remaining 2479 sources 
follow radio source redshift number density profile, we have shown one sample input to 
our simulations.  
}
\label{fig:redshift}
\end{figure}

In our analysis we need redshift distribution of JVAS/CLASS sources in order
to model the 
statistical alignments of radio polarization. Since the redshifts of
many sources are unknown, we adopt the 
following hybrid redshift model. We employ {\bf NASA/IPAC EXTRAGALACTIC  DATABASE (NED)
--}{\it`Retrieve Data for Near- Object/Position List'\footnote{\url{https://ned.ipac.caltech.edu/forms/nnd.html}}} 
tool and cross match all (above 1 mJy)  4400 JVAS/CLASS sources with NED 
objects 
and find nearby objects within radius 0.1 and 0.5 arcsec. We identify 1783 
sources with a search criteria of 0.1 arcsec and an additional 138 if
the larger radius of 0.5 arcsec is used. This leads to a total of
1921 sources with redshifts in 
NED catalogue. Most of these (1389) are quasars, a few (201) are listed as 
galaxies and only 11 are identified 
as radio sources. The remaining 320 sources are other types or unknown type. We show 
the redshift distribution of all these different classes of sources  
in Fig. \ref{fig:redshift}. Even so we do not find the redshift for 2479 (out of 4400) sources and we 
adopt the radio source redshift profile for these. This is justified because the redshift for radio sources is 
largely unknown and thus most likely the unknown redshift sources are radio only. Anyway, in our analysis we 
do not need precise redshifts, we only need redshift distribution to glean out the 
magnetic field statistical correlations. The radio redshift distribution is largely unknown and we only 
have a few small area deep survey observations to determine the redshift number density. 
We rely on  Combined EIS-NVSS Survey  of Radio Sources  (CENSORS) \citep{Best:2003,Rigby:2011} and  
the Hercules\citep{Waddington:2000,Waddington:2001} observed redshift number distribution and model the radial 
number density of remaining 2479 sources. This is the same redshift distribution as followed 
in \cite{Adi:2015nb,Tiwari:2015a} and presumably the best we can assume for JVAS/CLASS radio sources. 
We show the CENSORS and Hercules redshift number distribution fit \citep{Adi:2015nb} as {`fit'} in 
Fig. \ref{fig:redshift}, a sample redshift histogram, input to simulation, is also shown in the same figure.

We remove 187 sources from these 4400 sources as their angular positions 
coincide with other sources. We point out that the sources lie dominantly 
in the Northern hemisphere. Furthermore there are very few sources along the galactic plane. 
We use only this sample of 4213 sources for our analysis. 

\begin{figure}
\includegraphics[width=3.5in,angle=0]{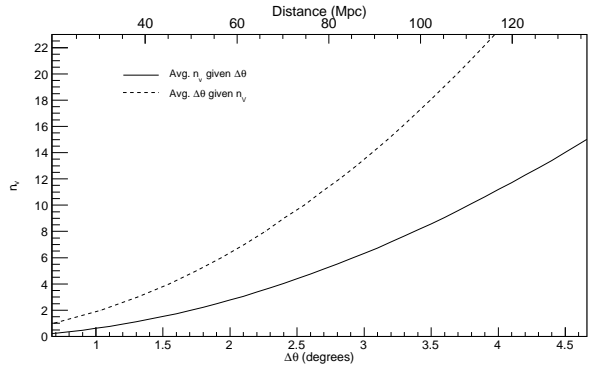}
  \caption{
The average angular separation for a fixed number of nearest neighbours ($n_v$) and vice versa.
Corresponding physical distance scale (Mpc) of the source 
assuming that it is located at $z=1$ is also given.  
}
\label{fig:nv_dis}
\end{figure}

The alignment statistic we use is defined for the number of nearest neighbours, $n_v$, or equivalently the angular   
separation $\Delta \theta$ between sources.  
For any chosen source $k$ we order the remaining sources in terms of
increasing angular separation from the source $k$. The closest 
$n_v$ sources, excluding the source $k$ itself, forms the required
nearest neighbour set. The minimum value of $n_v$ is clearly 1 and
the maximum value we explore is equal to 15 for reasons given later 
in section \ref{sc:results}.  
We can associate a mean $\Delta\theta$ with $n_v$ nearest neighbours 
by determining the $\Delta\theta$ corresponding to the nearest 
neighbour set of each source and taking the average over all sources.   
The mean value of $\Delta\theta$ increases monotonically with $n_v$. 
Hence $n_v$ also provides
a measure of the angular separation. 
In order to relate $n_v$ or $\Delta\theta$ to separation distance we assume that sources are located at mean redshift approximately equal to 1. 
 We need to make this assumption since we do not have 
redshifts for most of 
sources, indeed we only have the probability distribution. We have the redshifts for 1921 sources, most of them are 
quasars and are found to have larger distances scale correlations in $2D$ 
and $3D$ \citep{Pelgrims:2015}.
We show the average angular separation and the physical distance scale as a 
function of $n_v$ 
in Fig. \ref{fig:nv_dis}. The physical distance scale $L$ of a distant cluster
is related to its angular size $\Delta \theta$ by the formula 
$L=d_A\Delta\theta$, where $d_A$ is the angular diameter distance, 
computed at redshift $z=1$ using the standard Lambda Cold Dark Matter model \citep{Weinberg:2008}.
We clarify that \cite{PTiwari:2013pol} used the comoving distance instead of
the angular diameter distance. This leads to a change in our estimate of
the distance scale of alignment from 150 Mpc to 100 Mpc.


\section{Correlation Statistic}
\label{sc:SD}
In order to quantify the correlations between polarizations at different
positions we define two statistics, $S_D$ and $S'_D$. 
Consider the $n_v$ nearest neighbours of a source located at site $k$.
Let $\psi_i$ be the PA of the source at the $i^{th}$ site within the $n_v$ nearest neighbour set.
The dispersion of PAs in the neighbourhood of this site may be characterized by the measure,
\bea
d_{k} = \frac{1}{n_{v}} \sum_{i=1, i\neq k}^{n_{v}} \cos[2(\psi_{i}+\Delta_{i\rightarrow
k}) - 2\psi_{k})].
\label{eq:dispersion}
\eea
Here the factor $\Delta_{i\rightarrow k}$ arises since the polarizations
at two different points, labelled as $i$ and $k$,
on the celestial sphere have to be correlated
after making a
parallel transport from $i\rightarrow k$  \citep{Jain:2003sg}
 along the geodesic which connects the two positions.
The parameter $\psi_k$ in this equation is the PA at the site $k$.
The measure $d_k$ provides an estimate of the dispersion.
We point out that a large value of $d_k$ implies low  dispersion and vice versa.
The statistic is defined as \citep{Hutsemekers:1998,Jain:2003sg},
\bea
S_{D} = \frac{1}{n_{s}} \sum_{k=1}^{n_{s}} d_{k},
\label{eq:statistic}
\eea
where $n_s$ is the total number of data samples.
A strong alignment between polarization vectors implies a large value of 
$S_{D}$. We point out that the notation used in this work is different
from \cite{PTiwari:2013pol}. We have also dropped self correlations 
($i=k$) while calculating $d_{k}$ in equation (\ref{eq:dispersion}). This changes the
magnitude of $S_{D}$ significantly for small $n_v$  
but does not change the alignment significance results.    
However, as already mentioned above, in this paper we are considering the 
jackknife errors which 
are significantly larger than the variance of $S_D$ for the case of
 the random polarization samples, considered in 
our previous paper \citep{PTiwari:2013pol}. 
This does lead to an appreciable change in the significance of alignment.

Alternatively we define statistic  $S'_D$ where we measure the dispersion over a 
fixed angular separation ($\Delta \theta$) rather than measuring it over a fix number of nearest neighbours ($n_v$).
We point out that these two statistic $S_D$ and $S'_D$ will give same result 
for a spatially uniform source distribution since for a circle of a 
given radius  the nearest number of sources will be same everywhere. 
Hence, $S_D$ can be translated into 
$S'_D$ naively. However, the data sample is expected to show some deviation
from spatial uniformity and hence we expect small differences in the two 
measures, $S_D$ and $S'_D$.

\section{Procedure}
\label{sc:method}

We first determine the statistic $S_D$ ( see equation (\ref{eq:statistic}))
for the observed linear PAs. The same procedure is repeated for the theoretical 
PAs, which are assumed to be aligned perpendicular
to the background magnetic field. The magnetic field is generated by simulations
as explained in Section \ref{sc:PMF} and we glean out the galaxies 
according to redshifts details as 
given in Fig. \ref{fig:redshift}. For sources where we only have probability 
distribution of 
number density, we sample over 100 random outputs of redshifts generated
from the probability distribution shown in Fig. \ref{fig:redshift}. Finally, we calculate the 
$2D$ correlation statistics $S_D$ and $S'_D$.
The resulting theoretical values of $S_D$ are computed for a range of values of the spectral index 
$n_B$. The best fit value of $n_B$ is obtained by making a $\chi^2$ fit to the observed data.  

\begin{figure}
\includegraphics[width=4.0in,angle=0]{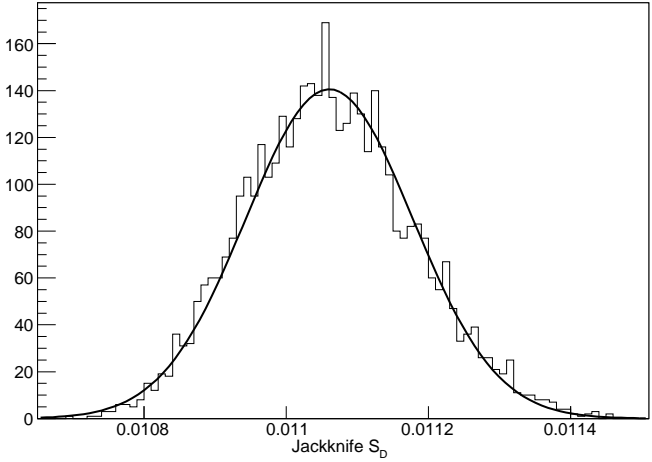}
\caption{ The distribution of jackknife sampled statistics $S_D$ for $n_v=10$. 
A fit to Gaussian is 
also shown. Note that the full sample statistics $S_D=0.01106$ is almost
equal to the jackknife sampled statistics $S_D$
distribution mean ($0.01106$)  and so the standard deviation in 
figure is roughly the $\delta S_D/ \sqrt{(N-1)}$.}
\label{fig:SD_dist}
\end{figure}

\begin{figure}
\includegraphics[width=5.0in,angle=0]{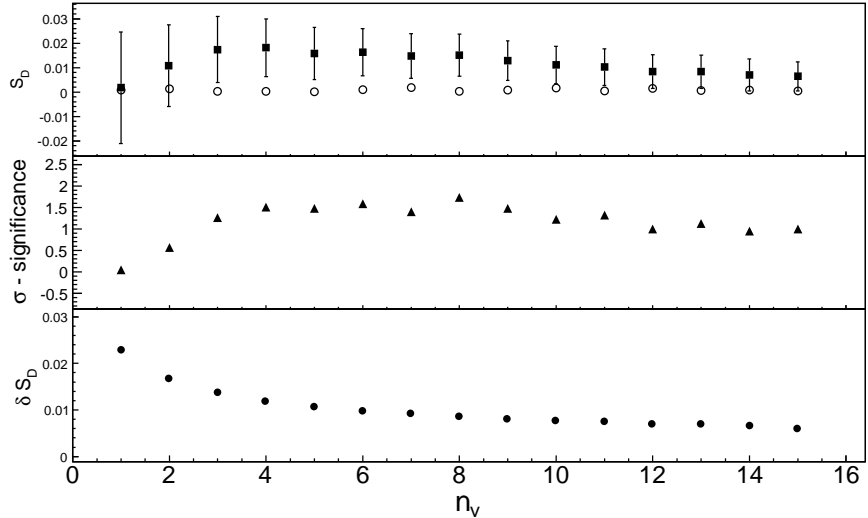}
  \caption{ The statistic $S_D$ for data with jackknife error bars along with  
 $S_D$ corresponding to randomly generated PAs (open circles) are shown in the top panel. 
The alignment 
significance considering jackknife errors is plotted in middle panel. For clarity we have 
plotted jackknife errors $\delta S_D$ for each $n_v$ in bottom panel.
}
\label{fig:SD}
\end{figure}

\begin{figure}
\includegraphics[width=5.0in,angle=0]{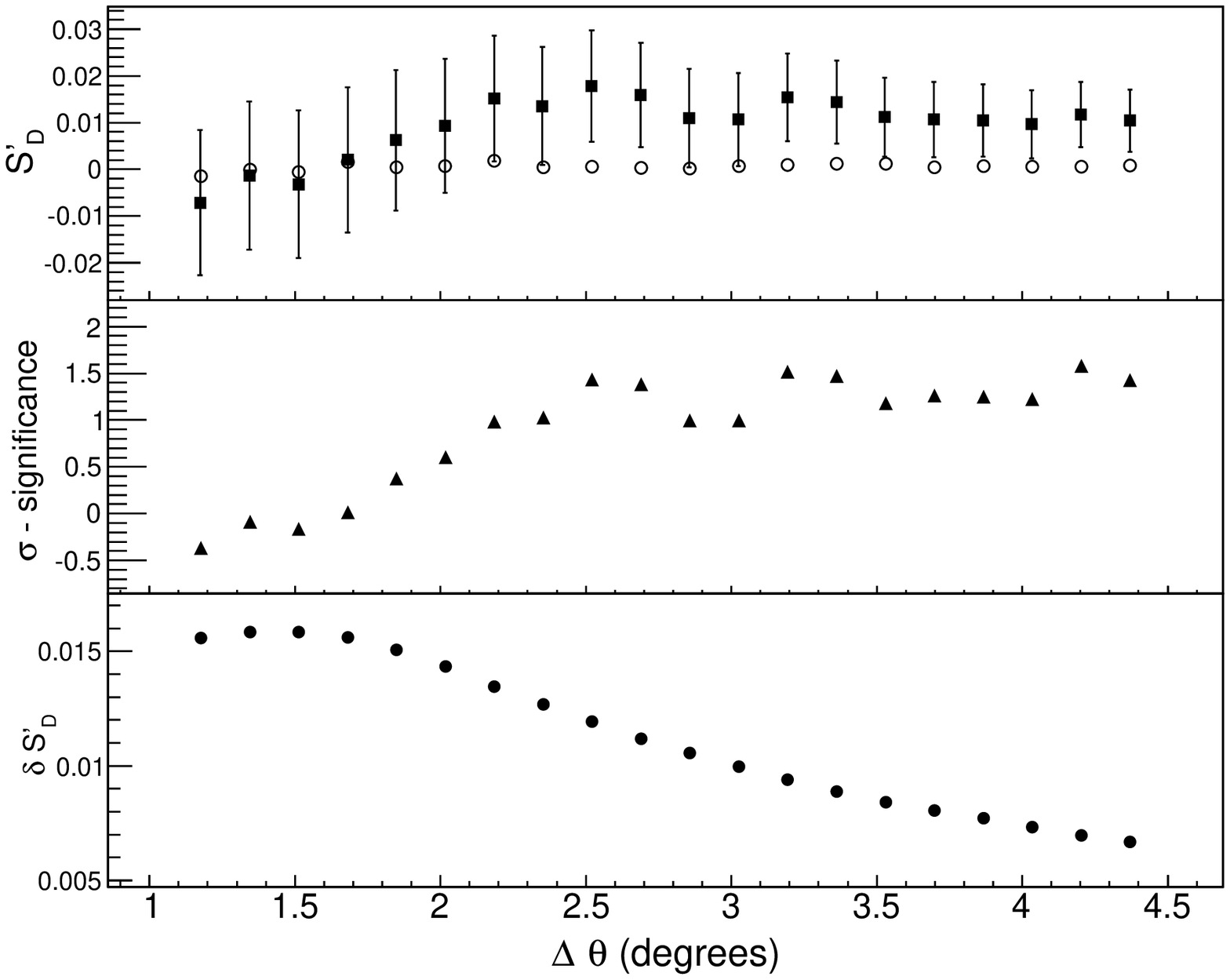}
  \caption{The statistics $S'_D$.  Data with jackknife error bars along with 
 $S'_D$ corresponding to random PAs (open circles) are shown in the top panel. 
The alignment         
significance considering jackknife errors is plotted in middle panel. The jackknife 
errors $\delta S'_D$ are in bottom panel.
}
\label{fig:SD2}
\end{figure}

In order to compute $\chi^2$ we need an estimate of the error in the
statistic $S_D$. We resort to jackknife errors as we do not know the exact errors in polarization measurements. 
For computing the 
jackknife errors, we resample the 
data by eliminating the $i^{{\rm th}}$ source and calculate the correlation statistics $S_D(i)$.  We repeat this process for 
all sources and calculate the $S_D(i)$ for $i=1$ to $4213$. We call the full sample statistics as $S_D$ and the 
jackknife error in its estimation is  given as 
\begin{equation}
(\delta S_D) ^2 = \frac{(N-1)}{N}\sum^{N}_{i=1} (S_D(i) -S_D)^2,  
\label{eq:deltaSD}
\end{equation}
 where $N$ is the total number of data point, which is 4213 in our case.  
We add that the jackknife errors do not include systematic errors. 
 The distribution of jackknife sampled $S_D$ and a Gaussian fit for $n_v =10$ 
is shown in Fig. \ref{fig:SD_dist}. Note that the mean of jackknife sampled statistics is almost the same as the 
full sample $S_D$ and so the standard deviation in Fig. \ref{fig:SD_dist} 
is roughly equal to $\delta S_D/ \sqrt{(N-1)}$. 
The $\delta S_D$ for $n_v =10$ as estimated from equation (\ref{eq:deltaSD}) is found to be $0.007716$. 
We similarly calculate the jackknife errors for all $n_v$ and angular scales $\Delta \theta$ 
for statistics $S_D$ and $S'_D$. In Fig. \ref{fig:SD} and \ref{fig:SD2} the jackknife errors are shown in 
lowest panel.

  
\section{Results}
\label{sc:results}
\begin{figure}
\includegraphics[width=4.0in,angle=0]{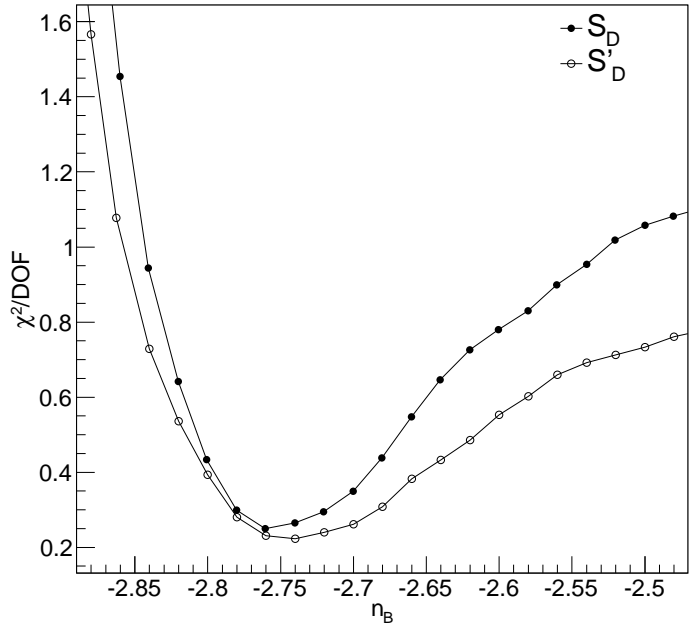}
  \caption{The $\chi^2/DOF$  as a function of the
 spectral index $n_B$ for 
the statistics $S_D$ and $S'_D$.  
}
\label{fig:Chinvdist}
\end{figure}

\begin{figure}
\includegraphics[width=4.0in,angle=0]{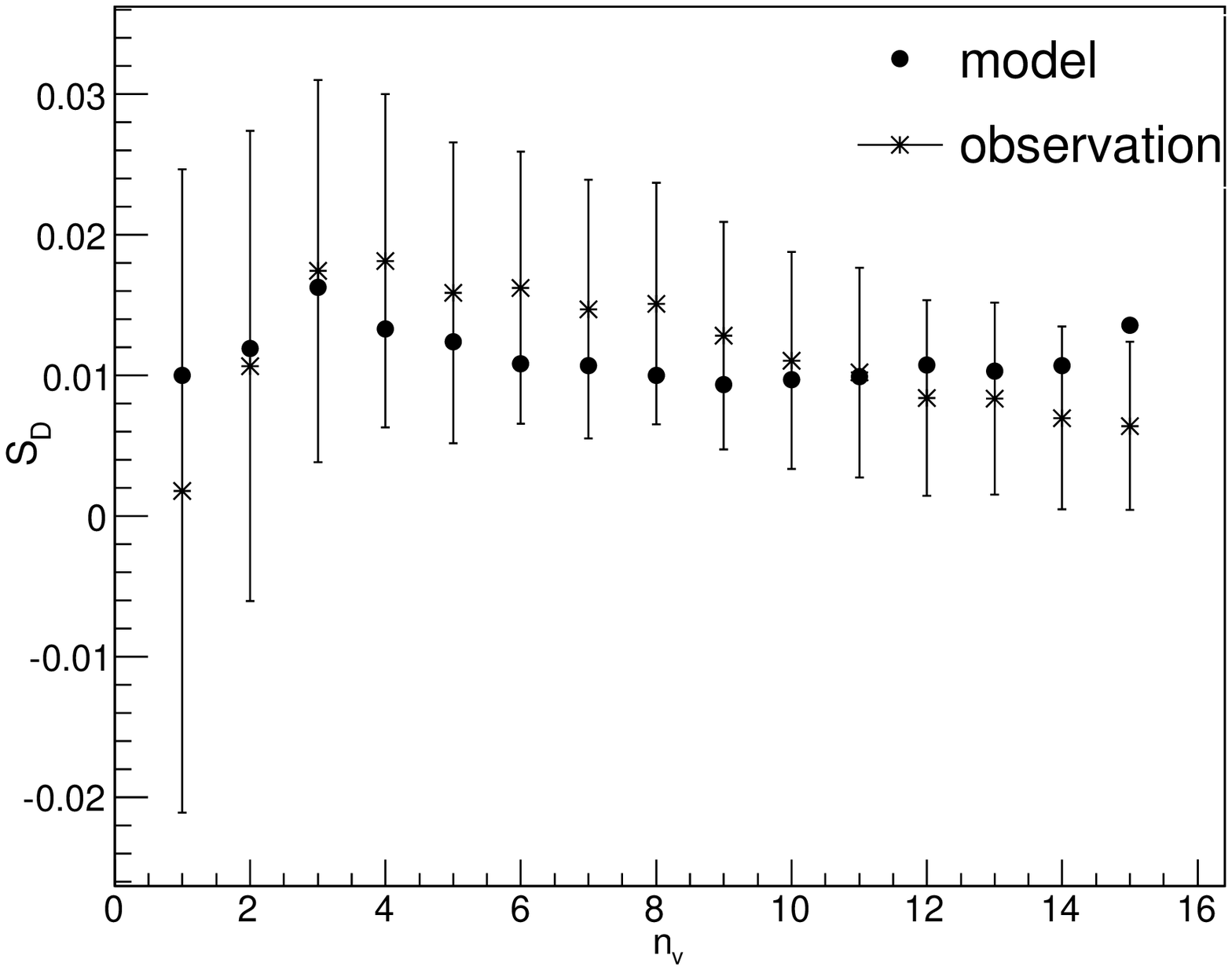}\\%
\includegraphics[width=4.0in,angle=0]{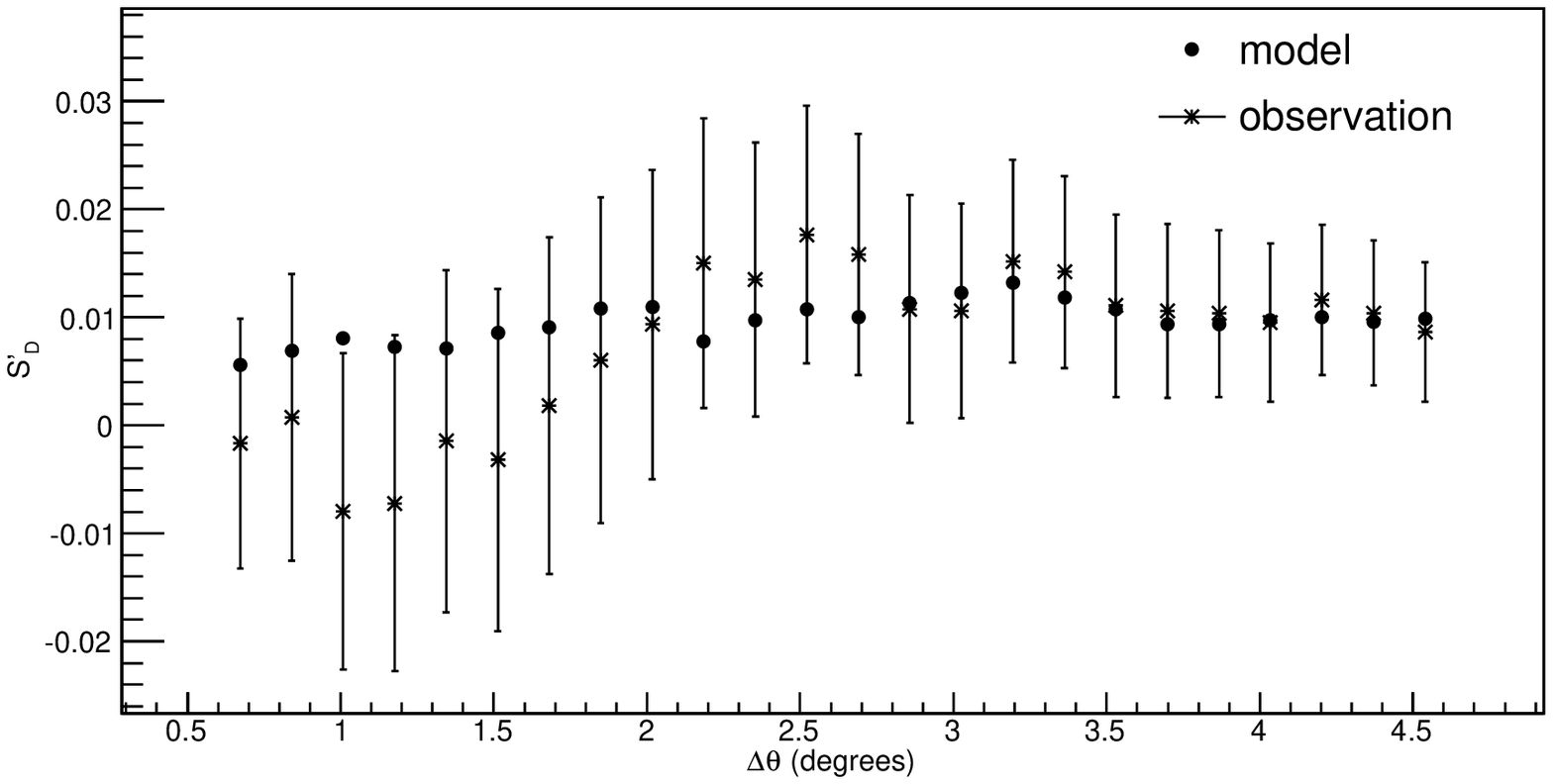}
\caption{The statistic $S_D$ (upper curve) and $S'_D$ (lower curve)
for the real data and simulated polarization from 
the background magnetic field at the source. This is for the best fitted spectral index 
$n_B=-2.76$ and $n_B=-2.74$ respectively.}
\label{fig:SD_comp}
\end{figure}

We simulate the $3D$ magnetic field for $n_B$ values in range $-2.20$ to $-2.98$, 
considering the power law spectrum as discussed in Section \ref{sc:PMF}. 
As described in Section \ref{sc:data}, we have redshifts of 1921 sources and 
for the remaining sources the redshift is generated assuming a fit obtained in \cite{Adi:2015nb}. 
We glean out the source location magnetic field directions from simulated magnetic field and 
calculate the model statistic $S_D$. We average over 100 realization to sample over fit 
generated random redshift positions. The data statistic $S_D$ along with alignment significance and 
jackknife errors is shown in Fig. \ref{fig:SD}.
The model $S_D$ values for different $n_B$ are fitted with observed data $S_D$.
The resulting $\chi^2$ is defined as
\beq
\chi^2=\sum_{n_{v}=1}^{15} \left(\frac{S_{D}(\rm {observed})-S_{D}(\rm {simulated})}{\delta S_D}\right)^2,
\label{eq:chi}
\eeq
where $\delta S_D$ is the jackknife estimate of error in $S_{D}(\rm {observed})$ 
(see Fig. \ref{fig:SD} and  \ref{fig:SD2}). 
We similarly  calculate $\chi^2$ for  $S'_{D}$. 
The maximum value of $n_v$ is set equal to 15. This choice is made since
the significance of alignment beyond $n_v=15$ is within 1-$\sigma$.
With one parameter the number of degrees of freedom (DOF) is equal to $14$. 
We present the $\chi^2/DOF$ values verses spectral index $n_B$ for statistic $S_D$ 
in Fig. \ref{fig:Chinvdist}. The minimum value of $\chi^2/DOF$ is found to be 
approximately  $0.25$ at $n_B =-2.76$. The data and best fitted simulated $S_D$ comparison 
for spectral index $n_B =-2.76$ is given in Fig. \ref{fig:SD_comp}. 
Including one sigma error, the extracted value of $n_B$ is found to be, $n_B=-2.76\pm 0.04$. We point out that 
the relatively low value of $\chi^2$ indicates that the jackknife errors are large. Nevertheless, even with 
such large errors we are able to clearly resolve the 
magnetic field spectral index as seen in Fig. \ref{fig:Chinvdist}.

The results for the case of the alternate statistic  $S'_D$ are 
also shown in Fig. \ref{fig:Chinvdist}. In this case we set the maximum value of
the fixed angular distance $\Delta \theta$ such as to include $15$ nearest neighbour as 
an average. This corresponds to $\Delta\theta=4.4^{\rm o}$. The full alignment results for statistic  $S'_D$  are shown in Fig. \ref{fig:SD2}. 
The angular distance corresponding to a mean value of $n_v$ is somewhat 
higher that the average 
distance for a fixed value of $n_v$. We have shown this difference in Fig. \ref{fig:nv_dis}. Despite this difference,
we obtain almost similar results for statistics $S'_D$. The 
best fit value of the spectral index is found to be 
$n_B=-2.74\pm 0.04$ (Fig. \ref{fig:Chinvdist}). The slight difference in the extracted
values of $n_B$ for the two statistics reflects the spatial 
non-uniformity of data. The value extracted using the statistic $S'_D$ 
may be more reliable since it includes sources within a fixed distance 
from a particular source rather than including a fixed number of nearest 
neighbours. The $\chi^2/DOF$ is slightly lower for $S'_D$ but the data points 
at low angular separation 
($\Delta \theta<2^{\circ}$) show large fluctuation due to  spatial non-uniformity of data. Nevertheless, 
the results from $S_D$ and  $S'_D$ agree within errors. 
It is interesting that the extracted value of $n_B$ is in good agreement with that obtained
by magneto-hydrodynamic simulations which suggest a spectral index of $-2.7$ \citep{Dolag:2002}.
The close agreement may be fortuitous but may be tested by future more
refined data.
In any case, theoretically, we do not expect a perfect agreement between 
these two indices. The polarization index ($n_B$), extracted from observations,
only acts as
a tracer of cluster magnetic field index ($n'_B$). If we assume that
the correlations in polarization are induced by those in the cluster 
magnetic field, we expect that the maximum level of alignment in the
polarization orientations would be equal to those of the background
magnetic field. The level of alignment, i.e. the value of the statistic
$S_D$ or $S'_D$, increases with $|n_B|$. Hence we expect that $|n_B|\le 
|n'_B|$.   

We also point out that since the best fit value of the spectral index 
is close to the theoretical expectations, we could not have expected 
a higher significance of alignment in this data set. 
If we 
compute the significance using the variance of alignment statistics of shuffled
PAs \citep{PTiwari:2013pol}, it is found to be 
 about 3 sigmas. However if we include the jackknife errors, the 
significance is at best about 1.5 sigmas, as shown in 
Figs. \ref{fig:SD} and \ref{fig:SD2}. In order to increase it to
3 sigmas we need to reduce the errors by a factor of two which requires
four times more data. Hence a clear test of our proposal can be made
by acquiring a suitably enhanced data set.

\section{Conclusion and Discussion}
\label{sc:conclusion}
We have presented a possible model to explain  the large scale radio
 polarization correlations. 
The model is based on two main assumptions that the galaxy jets are aligned with cluster 
magnetic field and that the jet orientations approximately mark the 
radio polarization
angles. We argue that the second assumption is well supported by data and
the cosmological magneto-hydrodynamic simulations provide some support for
the first assumption.  
We also find that our extracted value of the spectral index ($n_B=-2.74\pm 0.04$) is in good agreement with the value $-2.7$ obtained by magneto-hydrodynamic simulations. This is rather encouraging and provides additional support to  
our proposal. However given the inherent uncertainties in the
polarization data, we cannot 
claim this to be definitive evidence for our model, which requires 
 further testing with more refined data. 
 We conclude that the observed alignment in the JVAS/CLASS data 
can be successfully 
explained in terms of magnetic field correlations. The model can be 
further applied to other data sets 
and tested.  

\section*{Acknowledgements}
We thank  Ranieri Baldi and Noam Soker for discussion and advice.  
 We acknowledge CERN ROOT 5.27 for generating our plots. 
This work is supported in part at the Technion by a fellowship from the Lady Davis Foundation. 
\bibliographystyle{mn2e}
\bibliography{mf_radio}
\end{document}